\documentstyle[psbox]{elsart}

\begin{document}
\begin{frontmatter}
 \title{The X-ray CCD camera of the MAXI Experiment on the ISS/JEM}

 \author[Osaka,NASDA,CREST]{E. Miyata},
 \author[Osaka,NASDA,CREST]{H. Tsunemi},
 \author[Osaka]{H. Ogata},
 \author[Osaka]{D. Akutsu},
 \author[Osaka]{K. Yoshita},
 \author[Osaka]{Y. Hashimoto},
 \author[NASDA]{K. Torii},
 \author[RIKEN,NASDA]{M. Matsuoka},
 \author[RIKEN,NASDA]{N. Kawai},
 \author[RIKEN,NASDA]{A. Yoshida},
 \author[RIKEN,NASDA]{T. Mihara},
 \author[RIKEN]{T. Kotani},
 \author[RIKEN]{H. Negoro},
 \author[RIKEN]{H. Kubo},
 \author[RIKEN]{H. Matsumoto},
 \author[RIKEN]{Y. Shirasaki},
 \author[RIKEN]{B.C. Rubin},
 \author[RIKEN]{I. Sakurai}, and
 \author[Miyazaki]{M. Yamauchi}

 \address[Osaka]{Department of Earth \& Space Science, Graduate School 
 of Science, Osaka University, 1-1 Machikaneyama, Toyonaka, Osaka
 560-0043, Japan}
 \address[RIKEN]{The Institute of Physical and Chemical Research (RIKEN)}
 \address[NASDA]{Tsukuba Space Center, National Space Development Agency
 of Japan (NASDA)}
 \address[Miyazaki]{Faculty of Engineering, Miyazaki University}
 \address[CREST]{CREST, Japan Science and Technology Corporation (JST)}

 \begin{abstract}
  MAXI, Monitor of All-sky X-ray Image, is the X-ray observatory on the
  Japanese experimental module (JEM) Exposed Facility (EF) on the
  International Space Station (ISS). MAXI is a slit scanning camera
  which consists of two kinds of X-ray detectors: one is a
  one-dimensional position-sensitive proportional counter with a total
  area of $\sim$ 5000 cm$^2$, the Gas Slit Camera (GSC), and the other
  is an X-ray CCD array with a total area $\sim$ 200 cm$^2$, the
  Solid-state Slit Camera (SSC). The GSC subtends a field of view with
  an angular dimension of 1$^\circ\times 180^\circ$ while the SSC
  subtends a field of view with an angular dimension of 1$^\circ$ times
  a little less than 180$^\circ$. In the course of one station orbit,
  MAXI can scan almost the entire sky with a precision of 1$^\circ$ and
  with an X-ray energy range of 0.5$-$30~keV.

  We have developed the engineering model of CCD chips and the analogue
  electronics for the SSC.  The energy resolution of EM CCD for Mn K$\alpha$
  has a full-width at half maximum of $\simeq$ 182 eV. Readout noise is
  $\simeq$ 11 e$^-$rms.

  \end{abstract}

\end{frontmatter}

 \section{Introduction}

 The charge coupled device (CCD) is a standard X-ray detector due to its
 high X-ray detection efficiency, moderate X-ray resolving power, and
 high spatial resolving power. The Solid-state Imaging Spectrometer,
 SIS, onboard ASCA is the first CCD camera used as a photon counting
 detector and equipped on board the satellite (\cite{Tanaka}). Following
 SIS, many satellites such as HETE (\cite{Ricker}), Chandra
 (\cite{ACIS}), XMM (\cite{XMM}), and Astro-E (\cite{Hayashida}) carry a
 X-ray CCD camera on their focal planes.

 \section{MAXI}

 The International Space Station (ISS) will be placed in a nearly
 circular, high inclination (51.6$^\circ$), low Earth orbit having a 96
 minute orbital period with respect to a point in the sky. ISS will
 rotate synchronously with its orbit so that one side will always point
 toward the center of the Earth and the opposite side will permanently
 view the sky. A payload is attached to the main structure of the JEM
 which rotates and has unpredictable disturbances. Therefore, pointed
 observations are very difficult on the JEM. On the other hand,
 synchronous rotation with orbital revolution provides access to the
 entire sky in one orbit without a moving mechanism.  Considering these
 characteristics, we conclude that a monitoring mission or survey of a
 large field of the sky is suitable, and can produce significant
 scientific results.

\begin{figure}[ht]
   \begin{center}
    \psbox[xsize=.8#1]{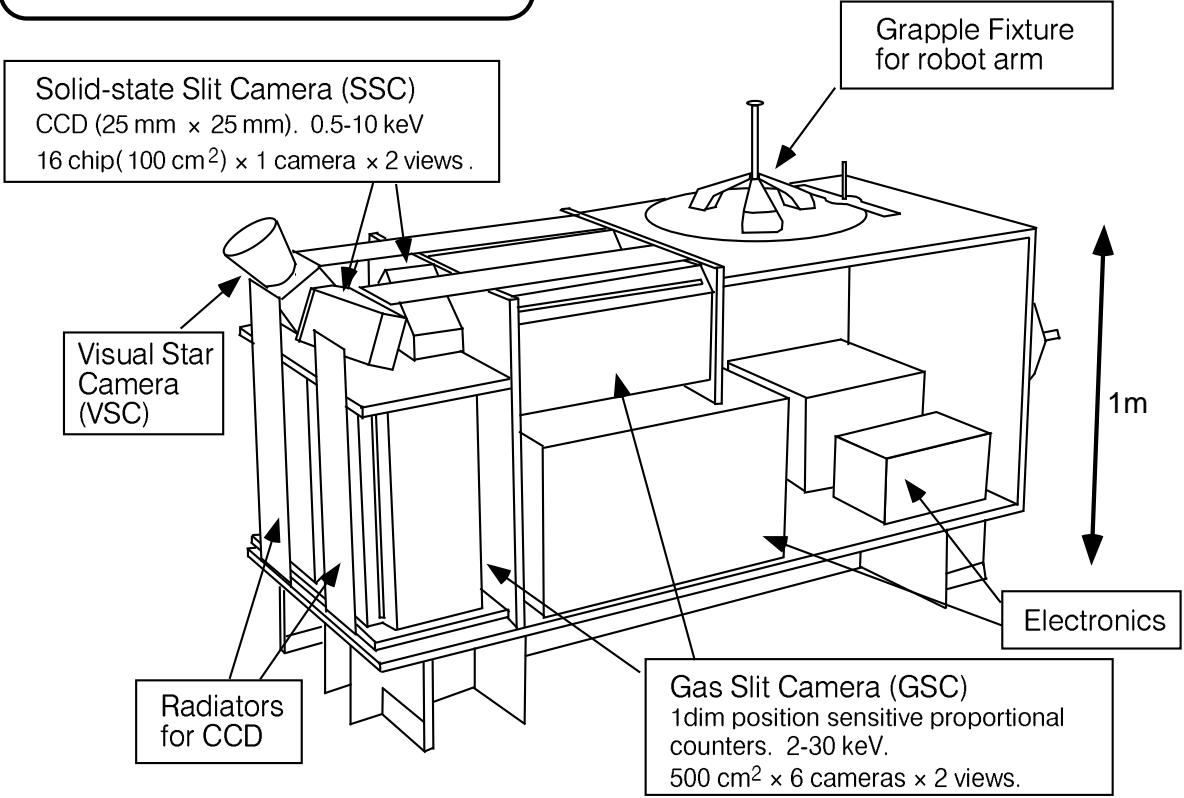}
    \caption{Schematic view of MAXI}
    \label{maxi}
   \end{center}
  \end{figure}

 The schematic view of MAXI is shown in figure~\ref{maxi}. MAXI can scan
 almost the entire sky with a precision of 1$^\circ$. MAXI carries two
 kinds of X-ray detectors: a one-dimensional position-sensitive
 proportional counter (GSC) and an X-ray CCD camera (SSC). Combining
 these two cameras, MAXI can monitor X-ray sources with an energy band
 of $0.5-30$ keV. The total weight of MAXI is about 500 kg.  Simulations
 of the data expected from MAXI have been performed in
 (\cite{Rubin}). A detailed description of MAXI can be found in
 (\cite{Matsuoka}, \cite{Matsuoka2}).

  \section{SSC}

\begin{figure}[t]
   \begin{center}
    \psbox[xsize=.6#1]{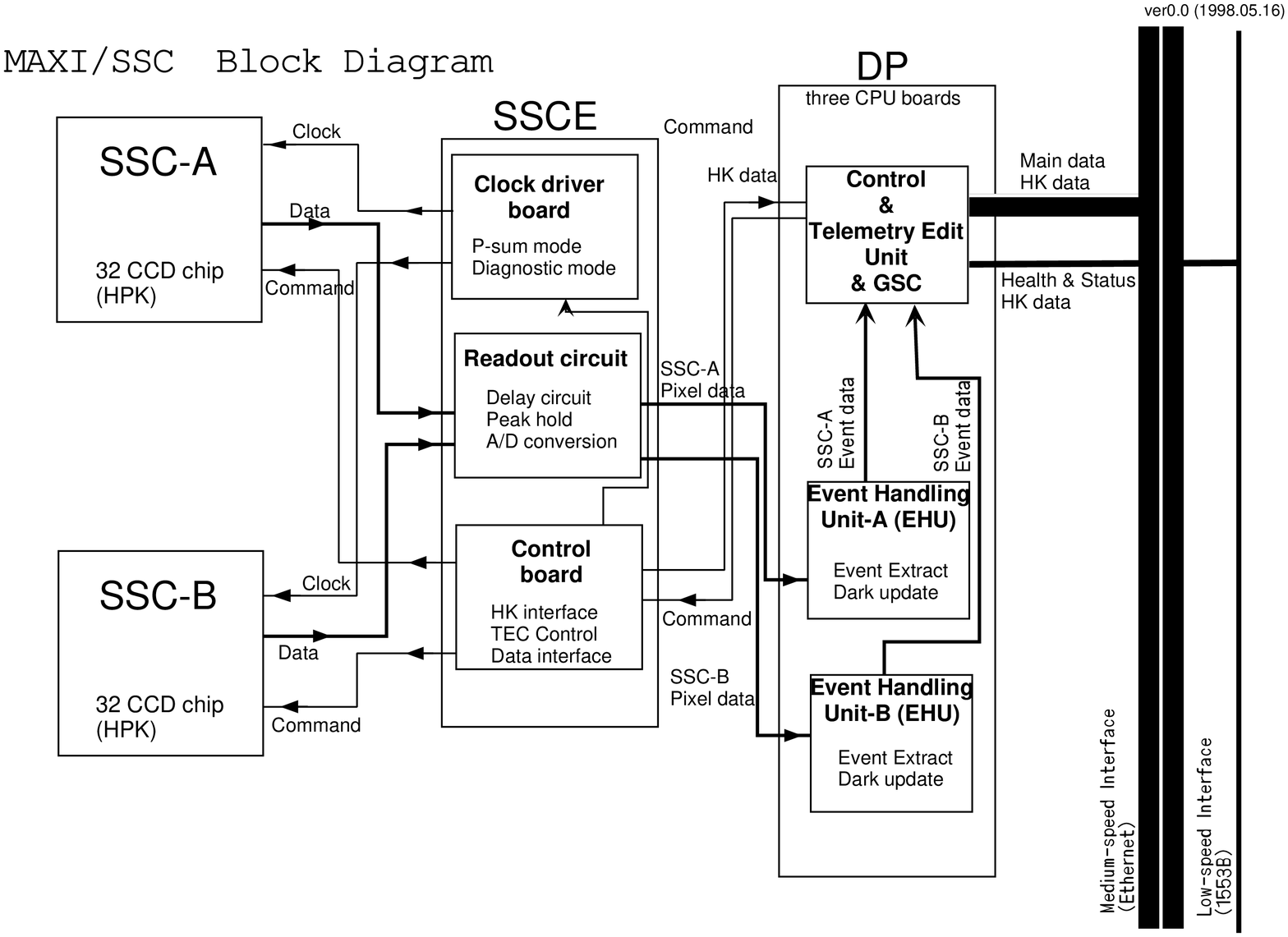}
    \caption{Block diagram of the SSC}
    \label{block-diagram}
   \end{center}
  \end{figure}

  The SSC is an X-ray CCD camera system. The SSC consists of two X-ray
  CCD cameras, each comprising 16 CCD chips.  The block diagram of the
  SSC camera is shown in figure~\ref{block-diagram}. The SSC consists of
  three parts: two CCD cameras, analogue electronics (SSCE), and a
  digital processing unit (DP). Detailed specifications of the SSC are
  shown in table~\ref{table:spec}.

  The CCD is fabricated by Hamamatsu Photonics K.K. (HPK). The CCD chip
  is three-side buttable with full-frame transfer and has 1024 $\times$
  1024 pixels of 24$\mu$m$\times$24$\mu$m size with two phase gate
  structures. The CCD chip is covered by $\sim$2000\AA\ Al to block
  optical light. The CCD is operated at $-60^\circ$C, which is
  achieved by using a passive cooling system and a Peltier cooling system
  (TEC). TEC is supported with glasses to hold out the launch shock.

  The SSCE is developed by Meisei Electronics. There are several CCD
  signal processing techniques (\cite{CCD} and references therein). To
  measure the voltage of each charge packet, we need a reference voltage
  between the floating level and the signal level. The correlated double
  sampling technique is widely used for this purpose.  In practice, it
  is advantageous to integrate or take the sum of the signals rather
  than merely spot sample floating and signal levels. Thus, a delay-line
  circuit is used in SIS/ASCA and an integrated circuit is introduced
  for SXC/HETE, ACIS/Chandra and XIS/Astro-E.  We plan to develop all
  these circuits for the SSC and will select the one that possesses
  the lowest readout noise.

  Since the data rate of CCD is fairly high, an onboard data reduction
  system is important. We developed an efficient reduction system based
  on our experiences with SIS/ASCA and XIS/Astro-E. There are three
  parts in DP: the control unit, the event handling unit (EHU), and the
  telemetry unit.  Two CPU boards (RAD 6000) on the VME bus will be used
  for EHU and another CPU board will be used for the control unit, the
  telemetry unit, and GSC data processing.

  There are two interfaces between MAXI and JEM Exposed Facility (EF):
  medium-speed interface (10Base-T ethernet) and low-speed interface
  (MIL1553B).  All CCD data will be downlinked through the ethernet
  whereas part of health and status (HS) data will be transferred
  through MIL1553B.

  Based on the SIS/ASCA, we have learnt much about radiation damage on
  the CCD (\cite{ayamashi}). One serious problem is the increase
  in dark current and its non-uniformity. To minimize the effects of
  radiation damage on the CCD, we allocate a dark level buffer for every
  pixel. The dark level for each pixel is updated for every frame based
  on the pulse height of pixel of interest. For the recovery of the
  radiation damage, we use an annealing process. However, we think that
  the radiation damage would be small because the lifetime of MAXI is
  two years (might be extended) and the orbit is lower than other
  missions ($\simeq$ 400km).

  Since the SSC is a one-dimensional X-ray camera, we use the spatial
  resolving power of the CCD only for the horizontal axis. Thus, we
  operate CCD in the parallel summing mode (same as the fast mode for
  SIS/ASCA). The vertical axis of the CCD corresponds to the time.  The
  binning number can be changed as $2^{\rm n}$ (n=2$\sim$8). 16 CCD
  chips in one camera are read cyclically using a multiplexer.

  \section{SSC Engineering Model}

\begin{figure}[t]
  \begin{center}
   \psbox[xsize=.8#1]{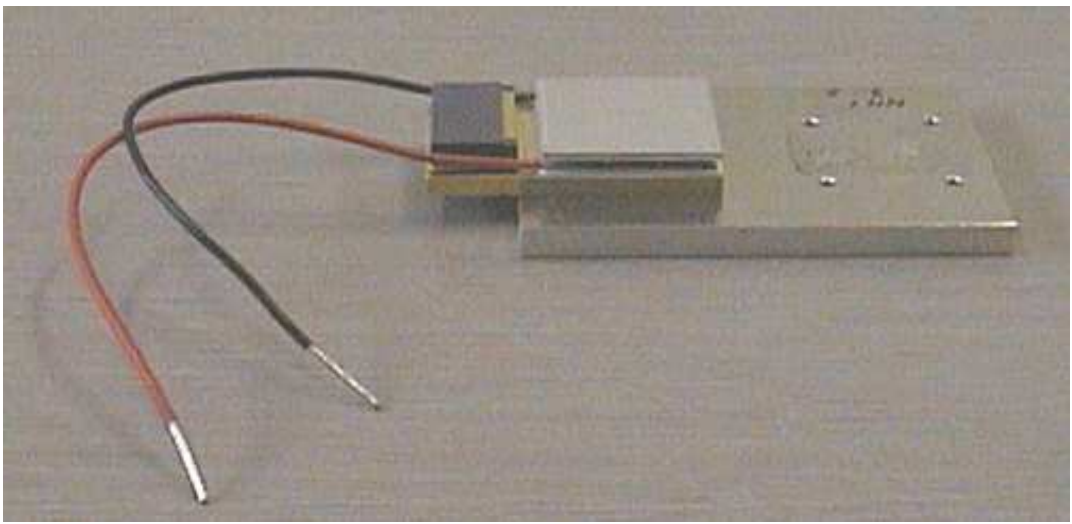}
   \caption{Photo of EM CCD chip}
   \label{ccd-chip}
  \end{center}
 \end{figure}

  The engineering model (EM) of the CCD chip has been completed and
  tested at the Osaka University X-ray CCD laboratory. EM of CCD is
  shown in figure~\ref{ccd-chip} where CCD is fixed on the Al plate. Two
  cables connected to the CCD are used for the Peltier cooler.

  There are three types of CCD produced for EM: a standard chip
  (standard), a deep depletion type I (deep-I), and a deep depletion
  type II (deep-II). There is a difference both in the depletion layer
  and in the dark current among these three types of CCDs. The details
  of these three chips can be referred to in Miyaguchi et al. (1999,
  \cite{Miyaguchi}).

  EM of the SSCE has been fabricated by MEISEI on the VME board. The
  function test of the EM SSCE is underway.

  \section{X-ray Responsivity}

  \subsection{Experimental Setup and Analysis}

  We evaluated the X-ray responsivity of deep-I EM CCD.  We cooled the
  CCD chip down to $-100^\circ$ with a He cryogenic system in the
  vacuum chamber. We used the C4880 CCD camera system, which is
  the X-ray CCD data acquisition system manufactured by HPK. Exposure
  time was set at 5 seconds.

  CCD frame files were transfered to a workstation through the ethernet
  with FITS format after they were acquired by C4880. HK information was
  collected with a workstation and stored in a hard disk.

  Dark current image was constructed with several CCD frame files using
  the same algorithm as that of XIS/Astro-E (\cite{Hayashida}). Before
  the X-ray event extraction, the dark current image was subtracted from
  each frame.

  \subsection{Results}

\begin{figure}[t]
   \begin{center}
    \psbox[xsize=.8#1]{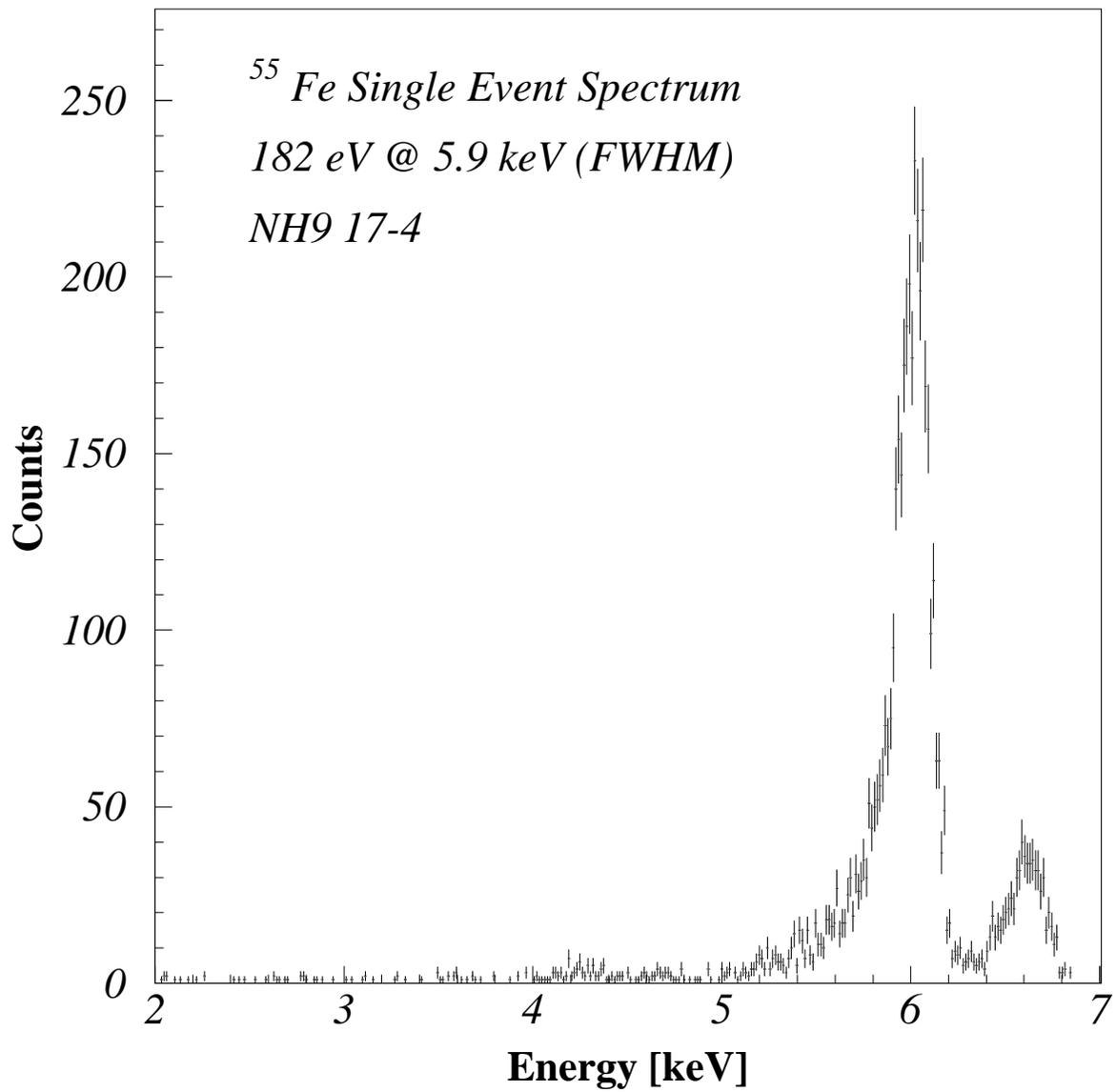}
    \caption{$^{55}$Fe spectrum obtained with the EM SSC camera}
    \label{fe-spec}
   \end{center}
  \end{figure}

  Figure~\ref{fe-spec} shows the energy spectrum of X-rays from
  $^{55}$Fe for single-pixel events. The split threshold is $\simeq$ 70
  eV.  Mn K$\alpha$ and K$\beta$ lines are clearly separated. The energy
  resolution of Mn K$\alpha$ has a full-width at half maximum of
  $\simeq$ 182 eV. Readout noise is $\simeq$ 11 e$^-$rms.

  Since the energy resolution of HPK CCD is $\sim$40\% less than that
  obtained by CCDs fabricated by the MIT Lincoln Laboratory
  (e.g. \cite{Hayashida}). HPK plans to improve the CCD to achieve
  performance comparable to those of other X-ray CCD devices.

  \section{Conclusion}

  MAXI is an X-ray all-sky monitor on the International Space Station
  and is due for flight in 2003. It is designed to scan almost the entire
  sky with a precision of 1$^\circ$ and with an X-ray energy range of
  0.5$-$30~keV in the course of one station orbit.

  We have developed the engineering model of the analogue electronics
  and the CCD chips for the X-ray CCD camera, SSC. We evaluated the X-ray
  responsivity of the EM CCD chip. The energy resolution of Mn K$\alpha$
  X-rays has a full-width at a half maximum of 182 eV. Based on the EM
  results, we will improve the performance of CCD and its electronics.

  \clearpage

  \begin{table}[htbp]
   \caption{Specifications of the SSC}
   \label{table:spec}
   \begin{tabular}{lc}
    \hline\hline
    CCD type & HPK CCD (Frame transfer; two phase) \\
    Number of cameras & 2 \\
    Number of CCD chips per camera & 16 \\
    Number of CCD pixels & 1024 (H) $\times$ 1024 (V) \\
    Pixel size & 24 $\mu$m $\times$ 24 $\mu$m \\
    Effective depletion depth (target) & $30-50 \mu$m \\
    Field of view & $\approx 1^\circ \times 180 ^\circ $ \\
    Angular resolution & $\approx 1 ^\circ$ \\
    CCD operating temperature (target) & $-60^\circ$C \\
    Annealing temperature & up to 20$^\circ$C \\
    Clocking mode & Normal (diagnostics) and P-sum (observation) \\
    \hline\hline
   \end{tabular}

  \end{table}

  \clearpage

  \clearpage

  \clearpage

  \clearpage

\end{document}